\documentclass[conference]{IEEEtran}
\usepackage{cite}
\usepackage{graphicx}
\graphicspath{}
\usepackage{multirow}
\usepackage{algorithm}
\usepackage{algorithmic}
\usepackage{url}

\usepackage{fancyhdr}
\usepackage{hyperref}
\fancypagestyle{firstpage}{% Page style for first page
  \fancyhf{}% Clear header/footer
  
  \fancyfoot[L]{{\em This paper was published in the Proceedings of the 2022 IEEE 16th International Conference on Semantic Computing (ICSC), pages 25-32, 2022. \copyright 2022 IEEE\\ Link to article abstract in IEEE Xplore: https://doi.org/10.1109/ICSC52841.2022.00012}}% Footer
}
\begin{document}

\pagenumbering{gobble}

\title{Impact of Stop Sets on Stopping Active Learning for Text Classification}

\author{
\IEEEauthorblockN{Luke Kurlandski}
\IEEEauthorblockA{Department of Computer Science\\
The College of New Jersey\\
Ewing, New Jersey 08618\\
Email: kurlanl1@tcnj.edu}
\and
\IEEEauthorblockN{Michael Bloodgood}
\IEEEauthorblockA{Department of Computer Science\\
The College of New Jersey\\
Ewing, New Jersey 08618\\
Email: mbloodgood@tcnj.edu}
}

\maketitle

\thispagestyle{firstpage}% firstpage page style for first page

%%%%%%%%%%%%%%%%%%%%%%%%%%%%%%%%%%%%%%%%%%%%%%
%%%%%%%%%%%%%%%%%%%%%%%%%%%%%%%%%%%%%%%%%%%%%%

\begin{abstract}
Active learning is an increasingly important branch of machine learning and a powerful technique for natural language processing. The main advantage of active learning is its potential to reduce the amount of labeled data needed to learn high-performing models. A vital aspect of an effective active learning algorithm is the determination of when to stop obtaining additional labeled data. Several leading state-of-the-art stopping methods use a stop set to help make this decision. However, there has been relatively less attention given to the choice of stop set than to the stopping algorithms that are applied on the stop set. Different choices of stop sets can lead to significant differences in stopping method performance. We investigate the impact of different stop set choices on different stopping methods. This paper shows the choice of the stop set can have a significant impact on the performance of stopping methods and the impact is different for stability-based methods from that on confidence-based methods. Furthermore, the unbiased representative stop sets suggested by original authors of methods work better than the systematically biased stop sets used in recently published work, and stopping methods based on stabilizing predictions have stronger performance than confidence-based stopping methods when unbiased representative stop sets are used. We provide the largest quantity of experimental results on the impact of stop sets to date. The findings are important for helping to illuminate the impact of this important aspect of stopping methods that has been under-considered in recently published work and that can have a large practical impact on the performance of stopping methods for important semantic computing applications such as technology assisted review and text classification more broadly.\end{abstract}

\IEEEpeerreviewmaketitle

%%%%%%%%%%%%%%%%%%%%%%%%%%%%%%%%%%%%%%%%%%%%%%
%%%%%%%%%%%%%%%%%%%%%%%%%%%%%%%%%%%%%%%%%%%%%%

\section{Introduction}
\label{sec:Introduction}

%%%%%%%%%%%%%%%%%%%%%%%%%%%%%%%%%%%%%%%%%%%%%%

Active learning (AL) is a widely used strategy for reducing data labeling costs. AL has been used for many semantic computing applications within natural language processing (NLP), such as text classification (TC) \cite{lewis1994, tong2001, schroder2020, yan2020, zhan2020, flores2021,bloodgood2018ICSC}, named entity recognition (NER) \cite{bloodgood2009CoNLL, cai2020}, and machine translation (MT)\cite{bloodgood2010ACL, miura2016, hazra2021}. Problem settings and recent developments in AL are discussed further in \cite{hino2020}. 

The essential concept behind AL is for the machine learning (ML) model to take an operational role in deciding which data instances should be labeled for learning. The active learning process begins with an unlabeled pool of data, an initially empty pool of labeled training data, and a batch size. In the first step of AL, the model queries an oracle (a human data labeler) for the labels of an initial batch of training data. Following this, the algorithm enters an iterative process: the model is trained on the existing set of labeled data; the model queries the oracle to label a new batch of training examples; and the labeled and unlabeled sets are adjusted accordingly. This process is shown in Algorithm~\ref{alg:ActiveLearning}, where $b$ refers to the batch size, $L$ the set of labeled examples, $U$ the unlabeled pool of examples, and $B$ the batch set of examples queried at each iteration of AL.

\begin{algorithm}
\caption{Active Learning Algorithm}
\begin{algorithmic}
\REQUIRE $U, b$ \COMMENT{input to algorithm} \\
\STATE $L  \gets $ select and annotate $b$ random examples from $U$\\
\STATE $U \gets U - L$
\REPEAT
\STATE Train model using $L$
\STATE $B \gets$ select and annotate $b$ examples from $U$ using the selection algorithm
\STATE $U \gets U - B$
\STATE $L \gets L \cup B$
\UNTIL{stopping criterion is met}
\end{algorithmic}
\label{alg:ActiveLearning}
\end{algorithm}

An essential component to any active learning setup is an effective stopping method that halts the data labeling process. Without a stopping method, the model would request annotations for every training example, defeating the purpose of AL. Past work has used the term ``aggressive" to characterize stopping methods that tend to prioritize saving annotations and the term ``conservative" to characterize stopping methods that tend to maximize model performance \cite{bloodgood2009CoNLL}. In general, it is desirable for stopping methods to be:
\begin{itemize}
	\item capable of reducing the number of labels required without sacrificing final performance,
	\item capable of adjustment to be more or less aggressive, and
	\item applicable and consistent across many AL settings, base learners, tasks, and datasets.
\end{itemize}

A simple stopping method would be to stop after an ad-hoc predetermined labeling budget has been exhausted. However, this runs a large risk of either obtaining a significantly lower-performing model or wasting large amounts of data labeling effort. 

Another simple approach would be to use a small set of labeled data to evaluate performance during learning and stop when performance levels off. However, previous work has found that not only does this method have the drawback of requiring additional labeled data, but also that this method doesn't work as well as state-of-the-art stopping methods that base the stopping decision on the result of clever measurements made on a large set of unlabeled data \cite{beatty2019ICSC}. Such a large set of unlabeled data has been termed a ``stop set"  \cite{bloodgood2009CoNLL}. 

Two of the most widely used classes of stopping methods are stabilizing predictions-based methods \cite{bloodgood2009CoNLL, bloodgood2013CoNLL, altschuler2019} and confidence-based methods \cite{vlachos2008}. 

A well-known confidence-based stopping method stops AL when model confidence on a large, representative stop set begins to decline \cite{vlachos2008}. We discuss confidence-based stopping methods in greater detail in section~\ref{sec:MethodsForStopping}.

Stabilizing Predictions (SP) stops AL when the agreement between successively trained models exceeds some threshold. SP computes the agreement between two models by comparing the predictions of each model on a large, representative, randomly selected set of examples that does not have to be labeled. We discuss the SP method in greater detail in section~\ref{sec:MethodsForStopping}. 

Two new stopping methods were recently proposed in \cite{mcdonald2020} and presented as having better performance than the leading existing stopping methods, however, the use of SP-stopping in those experiments did not use a proper stop set. Instead of a large unbiased representative stop set, the remaining unlabeled pool was used, which is a shrinking and systematically biased stop set. 

In this paper we investigate the impact of using the remaining unlabeled pool as a stop set and various randomly selected stop sets of different sizes. We show that the choice of stop set has a substantial impact on performance for both stability-based stopping and confidence-based stopping. Furthermore, the impact is different depending on whether the stopping method is a stability-based method or a confidence-based method.

In extensive experiments with multiple freely and publicly available widely used benchmark datasets, which will allow for others to reproduce our experiments or run further experiments, we find that when SP is used with an unbiased representative stop set as originally proposed in \cite{bloodgood2009CoNLL}, the performance of SP is stronger than that of the confidence-based methods, including those proposed in \cite{mcdonald2020}. The methods proposed in \cite{mcdonald2020} are similar to the stopping criterion proposed in \cite{vlachos2008}. We experimentally show the extent to which the method from \cite{vlachos2008} has more conservative performance than the methods proposed in \cite{mcdonald2020}. We find that the methods from \cite{mcdonald2020} and \cite{vlachos2008} perform better compared with themselves when using an unbiased representative stop set as originally proposed in \cite{vlachos2008}.

Our findings are important for helping to illuminate the impact of stop set choice on stopping method performance. We show experimentally that stop set choice can have a large practical impact on the performance of stopping methods for important semantic computing applications such as technology assisted review, and text classification more broadly. Contributions in this paper include:
\begin{enumerate}
	\item Illuminate the impact of the choice of using the remaining unlabeled pool as a stop set for SP in the experiments of \cite{mcdonald2020} and show how such a choice impacts SP stopping behavior. 
	\item Compare several stopping methods, using multiple stop sets with each stopping method, on several publicly available datasets to experimentally show how performance is affected by stop set choice. 
	\item Experimentally show the extent to which use of the biased remaining unlabeled pool as a stop set will tend to make stability-based methods stop more aggressively and make confidence-based methods stop more conservatively than when using an unbiased representative stop set. 
\end{enumerate}

Section~\ref{sec:RelatedWork} discusses related work, section~\ref{sec:MethodsForStopping} provides detailed descriptions of the stopping methods, section~\ref{sec:ExperimentalSetup} discusses our experimental setup, section~\ref{sec:ResultsAndAnalysis} presents our experimental results and analysis, and section~\ref{sec:Conclusion} concludes.

%%%%%%%%%%%%%%%%%%%%%%%%%%%%%%%%%%%%%%%%%%%%%%
%%%%%%%%%%%%%%%%%%%%%%%%%%%%%%%%%%%%%%%%%%%%%%

\section{Related Work}
\label{sec:RelatedWork}

Overall, given the large amount of research on stopping methods, there has been relatively less attention given to the impact of different stop sets than to the design and testing of the stopping method computations performed on the stop sets. To our knowledge there is no past work that focuses on investigating the impact of different stop sets on stopping method performance. There has been some consideration of ideas related to stop set impact as an auxiliary topic in several past works, which we explain here. In contrast to all of this past work, the current paper is focused on clarifying the importance and the impact of stop set selection on stopping method performance. 

A widely applicable stopping method based on model confidence is presented in \cite{vlachos2008}. The method is intended to be used in AL settings where uncertainty sampling is the query strategy and there is an expectation of a rise-peak-drop pattern in model confidence as AL proceeds. The method in \cite{vlachos2008} stops AL based on model confidence measured on a large separate representative set of examples. After each iteration of AL, this method runs the newly learned model over the large set of examples and estimates its confidence. When model confidence consistently drops on the large set of examples, this method indicates AL should stop. Although the term {\em stop set} is not used, in \cite{vlachos2008} there is some discussion of the set of examples that should be used for measuring model confidence and it is indicated that the set should be a separate large set of examples to minimize the risk of not being representative. There are little concrete details about the set of examples used in the experiments reported in \cite{vlachos2008}; it appears the test set might have been used, which would not be ideal as it could lead to overestimating the effectiveness of the stopping method. We experiment with this method using different stop sets and measure the impact of the different stop sets on performance. We refer to this method as Declining Confidence (DC). 

A widely applicable stopping method based on stabilizing predictions is presented in \cite{bloodgood2009CoNLL}. This is a particularly important method to investigate because it is widely applicable with few if any restrictions on when it can be used, and it has been found to work well in many settings \cite{bloodgood2009CoNLL, saha2015, burka2017, wiedemann2018, beatty2019ICSC, shmilovich2020, ren2021, pullar2021}. The method stops when successively trained models have high agreement in terms of their predictions on the examples in a large randomly selected set of examples. The authors call this set of examples a {\em stop set} \cite{bloodgood2009CoNLL}. In the main experiments, a large randomly selected fixed set of examples is used as the stop set. There is a small section of the paper that explores how stopping performance is impacted if the stop set is made smaller and it's found that if the set becomes too small the performance could begin to degrade, but it's concluded that the size required to be representative could vary depending on dataset and task so future work is needed to determine how large the stop set should be. We experiment with this method using different stop sets and measure the impact of the different stop sets on performance. We refer to this method as Stabilizing Predictions (SP). 

A theoretical analysis of SP stopping is presented in \cite{bloodgood2013CoNLL}. An equation for the variance in the agreement estimator between successive models is derived, which shows that the variance is inversely proportional to the size of the stop set. The variance estimator is suggested as a principled way to check stop set size in order to make sure reliable estimates of agreement can be obtained. 

Using labeled data versus using unlabeled data to inform the stopping decision is considered in \cite{beatty2019ICSC}. The tradeoff is that although the labels might give some extra information, that has to be balanced with the fact that using a stop set of labeled data for many applications will by necessity have to be kept relatively small whereas a stop set of unlabeled data can be large. It has been found that stopping methods that use unlabeled data in their decision are more effective than methods that use labeled data.

Two stopping methods are presented in \cite{mcdonald2020}. They are similar to the Declining Confidence (DC) method from \cite{vlachos2008}. The difference is that whereas the DC method requires a consistent drop in confidence, the methods from \cite{mcdonald2020} require a consistent period where confidence is not increasing, i.e., it could be staying the same or dropping. Another difference is that the methods were implemented with two different stop sets: the batch set, $B$, and the unlabeled pool, $U$, instead of a large representative set as suggested by \cite{vlachos2008}. We refer to these methods as Non-increasing Confidence (NC). 

In \cite{mcdonald2020} the performance of the NC methods are compared with current state-of-the-art methods, including SP, on a private technology assisted sensitivity review dataset. The conclusion reached in \cite{mcdonald2020} is that the leading method, SP, stops too aggressively on their private dataset and does not produce a model as well-learned as when using their NC methods. However, it is important to observe that SP was not implemented correctly in the experiments in \cite{mcdonald2020}. Instead of using a large unbiased representative stop set, the remaining unlabeled pool was used as the stop set. We hypothesize that this choice can cause SP to stop more aggressively and confidence-based methods such as DC and NC to stop more conservatively. In the current paper, we show that this is indeed the case and explain the reasons why. We compare the performance of NC with DC, which is not considered in \cite{mcdonald2020}, and quantify how much more conservative DC is on several datasets. We also compare the performance of NC with SP implemented in the normal way, which is not considered in \cite{mcdonald2020}, and find that then SP has compellingly stronger performance.

\section{Methods for Stopping}
\label{sec:MethodsForStopping}

In this section, we provide more details about the Stabilizing Predictions, Declining Confidence, and Non-increasing Confidence stopping methods, including details about their usage of stop sets. 
 
\subsection{Stabilizing Predictions}
\label{subsec:MethodsForStopping:StabilizingPredictions}

Stabilizing Predictions determines when to stop based upon the agreement between successively trained models \cite{bloodgood2009CoNLL}. The agreement is computed between the models' two sets of predictions on a large, randomly selected, fixed stop set of training examples. There are several ways the agreement can be computed. In \cite{bloodgood2009CoNLL} Cohen's Kappa statistic \cite{cohen1960} is suggested to account for agreement due to chance. 

We compute agreement in our implementation of SP as given by equation~\ref{eqn:Agreement}
\begin{equation}
\label{eqn:Agreement}
	agreement = \frac{A_o - A_e}{1 - A_e}
\end{equation}
where $A_o$ is the observed agreement between the two models and $A_e$ is the agreement expected between the two models by chance. We compute $A_o$ by the proportion of examples the two models agree upon. We compute $A_e$ as in equation~\ref{eqn:AgreementByChance}
\begin{equation}
\label{eqn:AgreementByChance}
	A_e = \sum_{k \in \{ +1, -1\}} P(k | c_1) P(k | c_2)
\end{equation}
where $c_i$ are the two models whose agreement is being computed and $P(k | c_i)$ is the probability that model $c_i$ labels an example as being in category $k$ \cite{bloodgood2009CoNLL}, \cite{cohen1960}.

The SP method determines when to stop based on whether the average of the agreements over a window of size $k$ exceeds a threshold $K$, where $k$ and $K$ are user-defined parameters. The SP method can be adjusted to be more aggressive by decreasing $k$ and $K$, and can be adjusted to be more conservative by increasing $k$ and $K$. Previous research has suggested that $k=3$ and $K=0.99$ work well as default values \cite{bloodgood2009CoNLL}.

The stop set of examples on which SP computes the agreement between successively trained models is an important part of the stopping method \cite{bloodgood2009CoNLL}, \cite{beatty2019ICSC}, \cite{bloodgood2013CoNLL}. Since SP uses this set of examples to make a generalization about the data distribution for the intended application of the learned model, the stop set should be an unbiased representation of the data. If it is systematically biased, the agreement between successively learned models on the elements in the stop set might not be indicative of the agreement between the learned models on application data as a whole. A larger stop set is more likely to be representative and unbiased, but a smaller set is more computationally efficient for computing agreement between the learned models. An easy way to select the stop set is to allocate a randomly chosen fixed percent of the data to it, such as $50\%$, as in \cite{beatty2019ICSC}. We denote such a randomly chosen unbiased stop set as $S$. 
Previous research found that the variance in the Kappa statistic \cite{bloodgood2013CoNLL} is inversely proportional to the size of the stop set. High variance in Kappa could cause SP to be behave unpredictably so the variance check from \cite{bloodgood2013CoNLL} is recommended as a principled way to ensure that the size is large enough to get low-variance estimates of agreement.

Recently, in \cite{mcdonald2020} SP was implemented using the remaining unlabeled pool, $U$, as the stop set. We will refer to this method as $\textrm{SP}_{U}$ and the original method as SP or $\textrm{SP}_{S}$. The choice of using the remaining unlabeled pool as the stop set is not explained in \cite{mcdonald2020}, providing evidence that the community could be under-appreciating the importance of stop set selection. When using the remaining unlabeled pool as the stop set, the set is changing at each iteration of AL. In particular, the size of this stop set is shrinking at each iteration of AL. Therefore, the number of examples in the remaining unlabeled pool at iteration $i-1$ and iteration $i$ will be different. Cohen's Kappa agreement needs to be computed on the same set of examples \cite{cohen1960}. When we implement $\textrm{SP}_{U}$, to ensure equally sized prediction lists, we compute the Kappa agreement between models trained at iteration $i-1$ and iteration $i$, but only use the examples in the unlabeled pool at iteration $i$. In contrast to $S$, $U$ is not representative of the application space, is systematically biased, and becomes increasingly more biased as AL proceeds. The extent and way that $U$ is biased will depend on the details of the AL setting, including the base learner that is used and the query strategy that is used. In \cite{mcdonald2020} a support vector machine (SVM) base learner is used with uncertainty sampling based on closest-to-hyperplane selection. This is known to result in a set of labeled data and a remaining set of unlabeled data that are biased \cite{bloodgood2009NAACL}. Intuitively, this makes sense because the algorithm is systematically selecting the examples that are the most difficult to classify to be labeled, leaving $U$ to consist increasingly of examples that are easier to classify as AL proceeds. We hypothesize that using this biased set as a stop set can be intuitively expected to yield higher agreement scores sooner than if an unbiased set were used. Our experimental results in section \ref{sec:ResultsAndAnalysis} confirm this hypothesis and help shed light on the extent of the impact.

%%%%%%%%%%%%%%%%%%%%%%%%%%%%%%%%%%%%%%%%%%%%%%

\subsection{Declining Confidence}
\label{subsec:MethodsForStopping:DecliningConfidence}

The Declining Confidence method is based on the premise that once all the informative examples in the unlabeled pool are labeled, further labeling could mislead the model and decrease its confidence \cite{vlachos2008}. DC determines when to stop AL based on when the model's confidence decreases for $\epsilon$ iterations where $\epsilon$ is a user-defined parameter. Previous research has suggested that using $\epsilon=3$ works well as a default. 

Given a stop set of training examples, $D$, DC measures the currently learned model's confidence by taking the mean of its confidence scores on every example in $D$. Using this notation, the DC method stops when the \textit{Conf} score, as in equation \ref{eqn:Conf}, decreases for $\epsilon$ consecutive iterations
\begin{equation}
\label{eqn:Conf}
	\textit{Conf}(D) = \frac{1}{|D|}\sum_{d \in D} confidence(d)
\end{equation}
where $confidence(d)$ is a measure of the model's confidence on datapoint $d$. In \cite{vlachos2008} when a support vector machine base learner is used, the confidence is measured by the distance of the datapoint $d$ from the separating hyperplane learned by the current model. 

Recently, in \cite{mcdonald2020}, two stopping methods were presented that are similar to DC from \cite{vlachos2008}. One difference is that the methods in \cite{mcdonald2020} use the batch set, $B$, and the remaining unlabeled pool, $U$, as their stop sets. In \cite{vlachos2008}, similarly to \cite{bloodgood2009CoNLL}, a large representative set is suggested to be used with DC. To investigate the impact of these different stop set choices, we implement DC using $B$, $U$, and a large randomly selected unbiased stop set $S$. We call these methods $\textrm{DC}_{B}$, $\textrm{DC}_{U}$, and $\textrm{DC}_{S}$, respectively\footnote{Observe that neither \cite{mcdonald2020} nor the current paper use $B$ as a stop set with SP because $B$ is a completely new set at each iteration of AL, and therefore, cannot be used with SP.}. Each method stops when $Conf(B)$, $Conf(U)$, or $Conf(S)$ decreases for $\epsilon=3$ iterations, respectively. We acknowledge that implementing DC with $B$ and $U$ is contrary to Vlachos' guidance about how to select the stop set because these sets are biased, changing, and in the case of $B$, small. We do so for illumination of the impact of stop set selection on the method since these sets were suggested for use as stop sets with the similar NC method in \cite{mcdonald2020}. 

\subsection{Non-increasing Confidence}
\label{subsec:MethodsForStopping:NonincreasingConfidence}

The two stopping methods presented in \cite{mcdonald2020}, TotalConf and LeastConf, both stop AL if the model's $Conf$ score (equation \ref{eqn:Conf}) measured over a stop set of training examples does not increase for $\epsilon$ iterations, where $\epsilon$ is a user-defined parameter. In \cite{mcdonald2020} the value of $\epsilon$ is set to three. Since the methods from \cite{mcdonald2020} stop when the model's confidence stops increasing, we refer to them as Non-increasing Confidence (NC). The difference between TotalConf and LeastConf is that TotalConf uses $U$ to compute confidence and LeastConf uses $B$. For brevity, we will call LeastConf $\textrm{NC}_{B}$ and TotalConf $\textrm{NC}_{U}$. 

We also implement NC with a large fixed unchanging unbiased representative stop set as recommended in \cite{bloodgood2009CoNLL,vlachos2008} and refer to this implementation of NC stopping as $\textrm{NC}_{S}$. We acknowledge that this was not suggested by the original authors in \cite{mcdonald2020}. We do so to illuminate the impact of stop set choice on NC stopping methods and we also hypothesize that using such an unbiased fixed stop set will strengthen the performance of NC stopping, which is confirmed experimentally in section~\ref{sec:ResultsAndAnalysis}.

%%%%%%%%%%%%%%%%%%%%%%%%%%%%%%%%%%%%%%%%%%%%%%
%%%%%%%%%%%%%%%%%%%%%%%%%%%%%%%%%%%%%%%%%%%%%%

\section{Experimental Setup}
\label{sec:ExperimentalSetup}

In our experiments, we perform text classification with seven widely used publicly available datasets. To make our results as reproducible as possible, we use the train-test split recognized by the community if such a split exists. Otherwise, we perform 10-fold cross validation. 

We use the ``bydate" version of the 20NewsGroups\footnote{\url{http://qwone.com/~jason/20Newsgroups/}} dataset and classify its newsgroup postings into 20 categories. We use the Reuters-21578 Distribution 1.0 ModApte split\footnote{\url{http://www.daviddlewis.com/resources/testcollections/reuters21578/}} as in \cite{joachims1998}, \cite{dumais1998} and classify the 10 largest categories of business news articles. We classify email spam from the SpamAssassin\footnote{\url{https://spamassassin.apache.org/old/publiccorpus/}} corpus and the ham25 version of the TREC\footnote{\url{https://plg.uwaterloo.ca/cgi-bin/cgiwrap/gvcormac/foo}} corpus. We use the results from the four largest categories of academic webpages in the WebKB\footnote{\url{http://www.cs.cmu.edu/afs/cs.cmu.edu/project/theo-20/www/data/}} dataset as in \cite{zhu2008a}, \cite{zhu2008b}, \cite{mccallum1998}, \cite{bloodgood2009CoNLL}. We use 10-fold cross validation for SpamAssassin, TREC, and WebKB; we macro average our results across folds and categories. 

In order to compare directly with the technology assisted sensitivity review experiment conducted in \cite{mcdonald2020} we tried to obtain the dataset that was used in that paper. Unfortunately, this dataset is private and cannot be shared with other researchers (personal communication with Graham McDonald). Instead, we perform TC on the publicly available RCV1-v2 dataset \cite{lewis2004} that has been used frequently in past works to simulate TAR \cite{cormack2015a}, \cite{cormack2015b}, \cite{cormack2016}, \cite{yang2019}, \cite{yang2021a}, \cite{yang2021b}. Most of these authors use a subset of the 103 RCV1-v2 categories; we use all 103 of them. RCV1-v2 is computationally demanding, so we do not use it in all of our experiments.

We use a support vector machine with a linear kernel and C=1, as done by \cite{mcdonald2020, bloodgood2009CoNLL, vlachos2008}. We use the closest-to-hyperplane selection algorithm as the query strategy during AL, as done by \cite{mcdonald2020, bloodgood2009CoNLL,vlachos2008}. We also use random selection, that is, passive learning, for a comparison experiment. We allocate $50\%$ of training examples to the randomly selected stop set, $S$, as in \cite{beatty2019ICSC}. We use a batch size of $0.5\%$, consistent with the findings of \cite{beatty2018ICSC}. We use a bag-of-words approach that only considers words that appear more than three times. For all datasets other than RCV1-v2\footnote{RCV1-v2 already comes with stop words removed.}, during preprocessing we remove 174 stop words using the \textit{Default English Stopwords List}\footnote{https://www.ranks.nl/stopwords}.

When comparing stopping methods, we report the number of annotations (i.e., labels) acquired (denoted by ANN in our tables), the proportion of annotations of the entire dataset acquired (ANN-P in tables), the learned model's F-1 measure (F1), F-2 measure (F2), accuracy (ACC), and balanced accuracy (BAC), all at the time the stopping method indicates to stop AL. These performance measures were used in \cite{mcdonald2020}, which will help to facilitate comparison. We also display these statistics when the model is given all possible training data (denoted by the column ``Final" in our tables). F-$\beta$ Measure is a weighted harmonic mean between precision and recall, where $\beta$ is a constant such that recall is $\beta$ times more important than precision. Accuracy is the proportion of the model's predictions that are correct. Balanced accuracy is the average of the recall scores for each class. We report these statistics for individual datasets, as well as a macro average across all datasets. In our tables we boldface entries that are the best-performing for each metric. Lower numbers are better for ANN and ANN-P because that reflects reducing the amount of labeled data required. Higher numbers are better for the other metrics since that reflects better system performance for end-users. 

%%%%%%%%%%%%%%%%%%%%%%%%%%%%%%%%%%%%%%%%%%%%%%
%%%%%%%%%%%%%%%%%%%%%%%%%%%%%%%%%%%%%%%%%%%%%%

\section{Results and Analysis} 
\label{sec:ResultsAndAnalysis}

%%%%%%%%%%%%%%%%%%%%%%%%%%%%%%%%%%%%%%%%%%%%%%

This section displays the results of our experiments and analyzes the results. Table \ref{tbl:Notation} shows the notation used throughout this section. 

\begin{table}[htbp]

\begin{center}
\scalebox{.9}{\begin{tabular}{| c | c | l |}

\hline
Symbol  & Type & Definition \\ 
\hline
{$\textrm{SP}$} & stopping method & {Stabilizing Predictions} \\
{$\textrm{DC}$} & stopping method & {Declining Confidence} \\
{$\textrm{NC}$} & stopping method & {Non-increasing Confidence} \\
{$S$} & stop set & {Standard fixed unbiased set of examples} \\
{$B$} & stop set & {Batch set of examples} \\
{$U$} & stop set & {Unlabeled pool of examples} \\
\hline
\end{tabular}
}
\end{center}
\caption{Notation used in this section. Stopping methods and stop set may be combined using subscript notation. For example, $\textrm{SP}_S$ indicates that the Stabilizing Predictions stopping method is used with a standard fixed unbiased stop set.}
\label{tbl:Notation}

\end{table}

\subsection{Impact of Stop Sets on Stabilizing Predictions} 
\label{subsec:ResultsAndAnalysis:ComparisonOfStabilizingPredictionsImplementations}

Table~\ref{tbl:StabilizingPredictionsImplementations} shows the performance of $\textrm{SP}_{S}$ and $\textrm{SP}_{U}$ for SVM-closest and SVM-random on all datasets (except RCV1-v2 for computational feasibility).

\begin{table}[htbp]

\begin{center}
\scalebox{.7}{\begin{tabular}{| l | l | l l l | l l l |}
\hline
{} & {} & \multicolumn{3}{ c |}{Active Learning} & \multicolumn{3}{c|}{Passive Learning} \\
\hline
{Dataset} &    {Stat} &   $\textrm{SP}_{U}$ &   $\textrm{SP}_{S}$ &     Final &           $\textrm{SP}_{U}$ &   $\textrm{SP}_{S}$ &     Final \\
\hline
20News	&      ANN 	&    \textbf{974} &   1088 &   11314 &           		3300 &   \textbf{2696} &   11314 \\
		&  ANN-P 	&  \textbf{0.086} &  0.096 &   1.000 &          		0.292 &  \textbf{0.238} &   1.000 \\
		&       F1 	&  0.714 &  \textbf{0.715} &   0.725 &          		\textbf{0.623} &  0.606 &   0.725 \\
		&       F2 	&  0.662 &  \textbf{0.664} &   0.681 &          		\textbf{0.558} &  0.539 &   0.681 \\
		&      ACC &  0.972 &  \textbf{0.973} &   0.973 &          		\textbf{0.964} &  0.963 &   0.973 \\
		&      BAC &  0.812 &  \textbf{0.814} &   0.824 &          		\textbf{0.757} &  0.748 &   0.824 \\
\hline
Reuters    	&      ANN &    \textbf{585} &    696 &    9655 &          			1190 &   \textbf{1084} &    9655 \\
		&  ANN-P 	&  \textbf{0.061} &  0.072 &   1.000 &          		0.123 &  \textbf{0.112} &   1.000 \\
		&       F1 	&  \textbf{0.786} &  0.782 &   0.761 &         		 \textbf{0.709} &  0.708 &   0.761 \\
		&       F2 	&  \textbf{0.746} &  0.738 &   0.717 &          		\textbf{0.654} &  \textbf{0.654} &   0.717 \\
		&      ACC &  0.968 &  \textbf{0.969} &   0.967 &         		 \textbf{0.961} &   0.96 &   0.967 \\
		&      BAC &  \textbf{0.856} &  0.851 &    0.840 &          		\textbf{0.806} &  \textbf{0.806} &    0.840 \\
\hline
Spam	&      ANN 	&    \textbf{618} &    655 &   10351 &          		  899 &    \textbf{863} &   10351 \\
Assassin	&  ANN-P 	&   \textbf{0.060} &  0.063 &    1.000 &          		0.087 &  \textbf{0.083} &   1.000 \\
		&       F1 	&  0.983 &  \textbf{0.985} &   0.997 &         		 0.956 &  \textbf{0.957} &   0.997 \\
		&       F2 	&  0.989 &  \textbf{0.991} &   0.997 &         		  \textbf{0.960} &   \textbf{0.960} &   0.997 \\
		&      ACC &  0.989 &  \textbf{0.991} &   0.998 &         		 \textbf{0.973} &  \textbf{0.973} &   0.998 \\
		&      BAC &   0.990 &  \textbf{0.992} &   0.998 &           		\textbf{0.970} &   \textbf{0.970} &   0.998 \\
\hline
TREC       &      ANN 	&   \textbf{5260} &   5565 &  152446 &           		8615 &   \textbf{8081} &  152446 \\
		&  ANN-P 	&  \textbf{0.034} &  0.036 &   0.999 &          		0.056 &  \textbf{0.053} &   0.999 \\
		&       F1 	&  0.988 &  \textbf{0.989} &    1.000 &        		  \textbf{0.971} &   0.97 &    1.000 \\
		&       F2 	&  \textbf{0.986} & \textbf{0.986} & 1.000 &    		 \textbf{0.967} &  0.966 &    1.000 \\
		&      ACC &  0.996 &  \textbf{0.997} &    1.000 &          		\textbf{0.991} &  \textbf{0.991} &    1.000 \\
		&      BAC &  0.991 &  \textbf{0.992} &   1.000 &           		\textbf{0.980} &  0.979 &      1.000 \\
\hline
WebKB     &      ANN &   \textbf{2059} &   2216 &   14176 &           		6707 &  \textbf{5268} &   14176 \\
		&  ANN-P 	&  \textbf{0.145} &  0.156 &   0.999 &          		0.473 &  \textbf{0.371} &   0.999 \\
		&       F1 	&  0.843 &  \textbf{0.856} &    0.980 &          		\textbf{0.817} &  0.781 &    0.98 \\
		&       F2 	&  0.812 &  \textbf{0.827} &   0.979 &         		 \textbf{0.814} &  0.777 &   0.979 \\
		&      ACC &  0.963 &  \textbf{0.966} &   0.995 &         		 \textbf{0.958} &  0.949 &   0.995 \\
		&      BAC &  0.889 &  \textbf{0.898} &   0.988 &        			  \textbf{0.894} &  0.873 &   0.988 \\
\hline
Avg          	&      ANN 	&   \textbf{1899} &   2044 &   39588 &          		 4142 &   \textbf{3598} &   39588 \\
		&  ANN-P 	&  \textbf{0.077} &  0.085 &   1.000 &          		0.206 &  \textbf{0.172} &    1.000 \\
		&       F1 	&  0.863 &  \textbf{0.865} &   0.893 &          		\textbf{0.815} &  0.804 &   0.893 \\
		&       F2 	&  0.839 &  \textbf{0.841} &   0.875 &         		 \textbf{0.791} &  0.779 &   0.875 \\
		&      ACC &  0.978 &  \textbf{0.979} &   0.987 &         		 \textbf{0.969} &  0.967 &   0.987 \\
		 &    BAC 	&  0.908 &  \textbf{0.909} &    0.930 &        		  \textbf{0.881} &  0.875 &    0.930 \\
\hline
\end{tabular}
}
\end{center}
\caption{The number of annotations and model performance when $\textrm{SP}_{U}$ and $\textrm{SP}_{S}$ stop for Active Learning and Passive Learning for all datasets (except RCV1 for computational feasibility). Bolded entries are the best-performing value for the relevant statistic.}
\label{tbl:StabilizingPredictionsImplementations}

\end{table}

As expected, active learning significantly outperforms passive learning, achieving similar performance with far fewer annotations. When the uncertainty sampling algorithm is used in active learning, we see that $\textrm{SP}_{U}$ is more aggressive than $\textrm{SP}_{S}$. We know from \cite{bloodgood2013CoNLL} that as the unlabeled pool shrinks, the variance in the Kappa statistic increases, and Stabilizing Predictions becomes more unstable. It was hypothesized in \cite{mcdonald2020} that the unlabeled pool becoming small in the later stages of AL could cause the more aggressive behavior of $\textrm{SP}_{U}$.

However, when examining the size of the unlabeled pool at the point when $\textrm{SP}_{U}$ stops AL, we see that $U$ is large at the stopping point for all datasets. At the stopping point, $U$ contains between $85\%$ and $96\%$ of the total dataset ($91\%$ on average). Since $U$ never gets small, the variance of the Kappa estimator never gets large. Therefore, the decreasing size of $U$ is not the main cause of the more aggressive stopping behavior of $\textrm{SP}_{U}$.

When an uncertainty sampling query strategy is used, $B$ and $U$ become biased sets. Since uncertainty sampling selects the most uncertain training examples for labeling, elements in $B$ are systematically biased such that they are difficult for the model to predict correctly. At each round of AL, the elements of $B$ are removed from $U$. This in turn causes the elements in $U$ to be biased such that they are easier for the model to predict correctly. Since the examples in $U$ are easier to predict, the consecutively trained models during AL will be more likely to agree on the predictions of the examples in $U$. This will make the Kappa agreement exceed its threshold sooner, and cause $\textrm{SP}_{U}$ to stop more aggressively. 

When a random sampling algorithm is used, that is, passive learning, $U$ should be representative of the dataset except that it is by definition not permitted to have any representation in the training data. This will make $U$ slightly harder to achieve agreement on when compared with an unbiased predetermined random set that is not artificially prevented from having representation in the training data. Therefore, we would expect $\textrm{SP}_{U}$ to behave more conservatively during passive learning than $\textrm{SP}_{S}$. Table~\ref{tbl:StabilizingPredictionsImplementations} confirms this, where we see that $\textrm{SP}_{U}$ is more conservative than $\textrm{SP}_{S}$ during passive learning. 

$\textrm{SP}_{U}$ requires a non-standard approach to compute Cohen's Kappa agreement between successively trained models, was found to behave over-aggressively by \cite{mcdonald2020} during AL, and uses a biased non-representative stop set. For these reasons, we suggest referring to the original authors of SP and implementing the method with a fixed unbiased representative stop set $S$. We compare $\textrm{SP}_{S}$ with the confidence-based stopping methods in the next section.

%%%%%%%%%%%%%%%%%%%%%%%%%%%%%%%%%%%%%%%%%%%%%%

\subsection{Impact of Stop Sets on Confidence Based Stopping Methods and Comparison with the SP Method}
\label{subsec:ResultsAndAnalysis:StopSetsAndStoppingMethods}

In table~\ref{tbl:StopSetsAndStoppingMethods}, we compare the stopping results from Declining Confidence and Non-increasing Confidence using the batch set, the unlabeled pool, and a large, randomly selected, fixed set as stop sets during AL. We also add the performance of $\textrm{SP}_{S}$ as a column for ease of comparison of stabilizing predictions stopping performance with confidence-based stopping performance. 

\begin{table}[htbp]

\begin{center}
\scalebox{.7}{\begin{tabular}{|l|l|lllllll|l|}

\hline
{Dataset}  & Stat &  $\textrm{SP}_{S}$ &   $\textrm{DC}_{B}$ &   $\textrm{DC}_{U}$ &   $\textrm{DC}_{S}$ &   $\textrm{NC}_{B}$ &   $\textrm{NC}_{U}$ &   $\textrm{NC}_{S}$ &    Final \\ 

\hline
20News		& ANN      	&   1088 &   3395  	&   {5793} &   1707 &    \textbf{265} &   1567 &    567  &  11314 \\
			& ANN-P 	&  0.096 &    0.300 	&  0.512 	&  0.151 &  \textbf{0.023} &  0.139 &   0.050 &      1.000 \\
			& F1       	&  \textbf{0.715} 	& 0.577 	&  0.608 & 0.598 & {0.488} &  0.539 &   0.530 &  0.725 \\ 
			& F2      	&  \textbf{0.664} &   0.520 &  0.553 &  0.544 &  0.419 &  0.476 &   0.470 &  0.681 \\
			& ACC      &  \textbf{0.973} &  0.962 &  0.963 &  0.962 &  0.956 &  0.958 &  0.957 &  0.973 \\
			& BAC      &  \textbf{0.814} &   0.740 &  0.757 &  0.752 &  0.688 &  0.717 &  0.714 &  0.824 \\
\hline		
RCV1		& ANN      		&    \textbf{6497} & 	60375 & 	67547 & 	14422 & 	20946 & 	46463 & 	8787 & 	   73697 \\
			& ANN-P	  	&    \textbf{0.088} &     0.819 &     0.917 &     0.196 &     0.284 &     0.630 &    0.119 &     1.000 \\
			& F1       		&    \textbf{0.525} &     0.520 &     0.515 &     0.522 &     0.508 &     0.516 &    0.518 &     0.515 \\
			& F2      		&    0.490 &      \textbf{0.502} &     0.499 &     0.491 &     0.479 &     0.493 &    0.484 &     0.500 \\
			& ACC     		&     \textbf{0.982} &     0.979 &     0.979 &     0.981 &     0.981 &     0.980 &    0.981 &     0.979 \\
			& BAC     		&    0.725 &      \textbf{0.736} &     0.735 &     0.728 &     0.723 &     0.730 &    0.725 &     0.736 \\
\hline
Reuters		& ANN     	&  \textbf{696}    &   6955 &   {8713} &  1785 &    758 &  4202 &    720 &   9655 \\
			& ANN-P 	&  \textbf{0.072} &   0.720 &  0.902 &  0.185 &  0.079 &  0.435 &  0.075 &      1.000 \\
			& F1       	&  \textbf{0.782} &  0.711 &  0.746 &  0.761 & {0.683} &  0.706 &  0.706 &  0.761 \\
			& F2      	&  \textbf{0.738} &  0.663 &  0.703 &  0.714 &  0.622 &  0.656 &  0.656 &  0.717 \\
			& ACC     	&  \textbf{0.969} &  0.963 &  0.965 &  0.967 &   0.960 &   0.960 &   0.960 &  0.967 \\
			& BAC     	&  \textbf{0.851} &  0.813 &  0.833 &  0.839 &  0.789 &  0.807 &  0.808 &   0.840 \\
\hline
Spam		& ANN      	&  \textbf{655} &  {10337} & {10337} &   2111 &   4579 &  {10337} & 1424 &  10351 \\
Assassin		& ANN-P	&  \textbf{0.063} &  0.999 &  0.999 &  0.204 &  0.442 &  0.999 &  0.138 &      1.000 \\
			& F1       	&  {0.985} &  \textbf{0.999} &  \textbf{0.999} &  0.994 &  0.996 &  \textbf{0.999} &  0.993 &  0.997 \\
			& F2      	&  0.991 &  \textbf{0.999} &  \textbf{0.999} &  0.997 &  0.998 &  \textbf{0.999} &  0.996 &  0.997 \\
			& ACC     	&  0.991 &  \textbf{0.999} &  \textbf{0.999} &  0.996 &  0.997 &  \textbf{0.999} &  0.995 &  0.998 \\
			& BAC     	&  0.992 &  \textbf{0.999} &  \textbf{0.999} &  0.997 &  0.998 &  \textbf{0.999} &  0.996 &  0.998 \\
\hline
TREC 		& ANN      	&   \textbf{5565} &  143103 & {152478} &  13113 &  113218 & 22716 & 12808 &  152531 \\
			& ANN-P 	&  \textbf{0.036} &   0.938 &       1.000 &  0.086 &   0.742 &  0.149 &  0.084 &   0.999 \\
			& F1       	&  {0.989} &       \textbf{1.000} &       \textbf{1.000} &  0.994 &   0.998 &  0.995 &  0.994 &       1.000 \\
			& F2      	&  0.986 &   0.999 &       \textbf{1.000} &  0.992 &   0.998 &  0.993 &  0.992 &       1.000 \\
			& ACC     	&  0.997 &       \textbf{1.000} &   \textbf{1.000} &  0.998 &   0.999 &  0.998 &  0.998 &       1.000 \\
			& BAC    	&  0.992 &       \textbf{1.000} &   \textbf{1.000} &  0.995 &   0.999 &  0.996 &  0.995 &       1.000 \\
\hline
WebKB		& ANN     	&   2216 &   9981 &  {11753} 			&  3846 &    \textbf{958} 	& 6651 &   2241 		&  14176 \\
			& ANN-P	&  0.156 &  0.703 &  0.828 			&  0.271 &  \textbf{0.068} 	&  0.469 &  0.158 &  0.999 \\
			& F1       	&  0.856 &   0.870  &   \textbf{0.940} 		&  0.893 &  {0.588} 		&  0.825 		&  0.784 &   0.980 \\
			& F2      	&  0.827 &  0.855 &  \textbf{0.929} 		&  0.871 &  0.546 		&  0.798 &   0.750 &  0.979 \\
			& ACC     	&  0.966 &  0.977 &  \textbf{0.989} 		&  0.978 &  0.917 		&  0.964 &  0.954 &  0.995 \\
			& BAC     	&  0.898 &  0.919 &   \textbf{0.960} 		&  0.925 &  0.746 		&  0.884 &  0.857 &  0.988 \\
\hline
Avg			& ANN      		&     \textbf{2786} &  39024 &  42770 &  6164 &  23454 &  15323 &  4425 &  45273 \\
			& ANN-P  		&    \textbf{0.085} &    0.747 &     0.860 &   0.182 &    0.273 &     0.470 &   0.104 &      1.000 \\
			& F1       		&    \textbf{0.809} &     0.78 &    0.801 &   0.794 &     0.71 &    0.763 &   0.754 &     0.830 \\
			& F2       		&    \textbf{0.783} &    0.756 &     0.780 &   0.768 &    0.677 &    0.736 &   0.725 &    0.812 \\
			& ACC     		&   0.980 &     0.980 &     \textbf{0.982} &    0.980 &    0.968 &    0.976 &   0.974 &    0.985 \\
			& BAC      		&   0.879 &    0.868 &     \textbf{0.881} &   0.873 &    0.824 &    0.856 &   0.849 &    0.898 \\
\hline
\end{tabular}
}
\end{center}
\caption{The number of annotations and model performance when a stopping method stops. Bolded entries are the best-performing value for the relevant statistic.}
\label{tbl:StopSetsAndStoppingMethods}

\end{table}

Table~\ref{tbl:StopSetsAndStoppingMethods} shows that an unbiased stop set, $S$, is the best stop set to use with DC. Table~\ref{tbl:StopSetsAndStoppingMethods} shows that when implemented with $U$ and $B$, DC uses far more annotations than is required and does not achieve any meaningful performance boost compared to $\textrm{DC}_{S}$. Comparing $\textrm{DC}_{S}$ with SP, from the average row, we see that $\textrm{DC}_{S}$ uses more than twice as many annotations as SP and all performance measures except ACC are worse than those of SP.

For NC, we see that the methods presented in \cite{mcdonald2020} have very different behavior on different datasets. For the individual datasets in table~\ref{tbl:StopSetsAndStoppingMethods}, we see that $\textrm{NC}_{B}$ does not learn an effective model for 20NewsGroups, Reuters, and WebKB, while it requests wastefully large numbers of annotations for RCV1, SpamAssassin, and TREC. $\textrm{NC}_{U}$ does not learn a good model for 20NewGroups and requests many more annotations than are required for all other datasets. In both the individual dataset results and the overall average results, we see that NC performs best when implemented with a large representative stop set, $S$. This strategy was first implemented in this work and was not considered in \cite{mcdonald2020}, yet it appears to be the highest-performing variant of NC. In the average row, we see that $\textrm{NC}_{S}$ attains essentially the same model performance as $\textrm{DC}_{S}$, but uses about half of the annotations. Therefore, it is reasonable to conclude that $\textrm{NC}_{S}$ is the best out of all the confidence-based methods. However, its performance is still not as strong as SP. The average row shows that $\textrm{SP}_{S}$ requires less labeled data and produces higher-performing models across all performance measures than $\textrm{NC}_{S}$. 

\begin{figure}[htbp]
\centerline{\includegraphics[scale=.6]{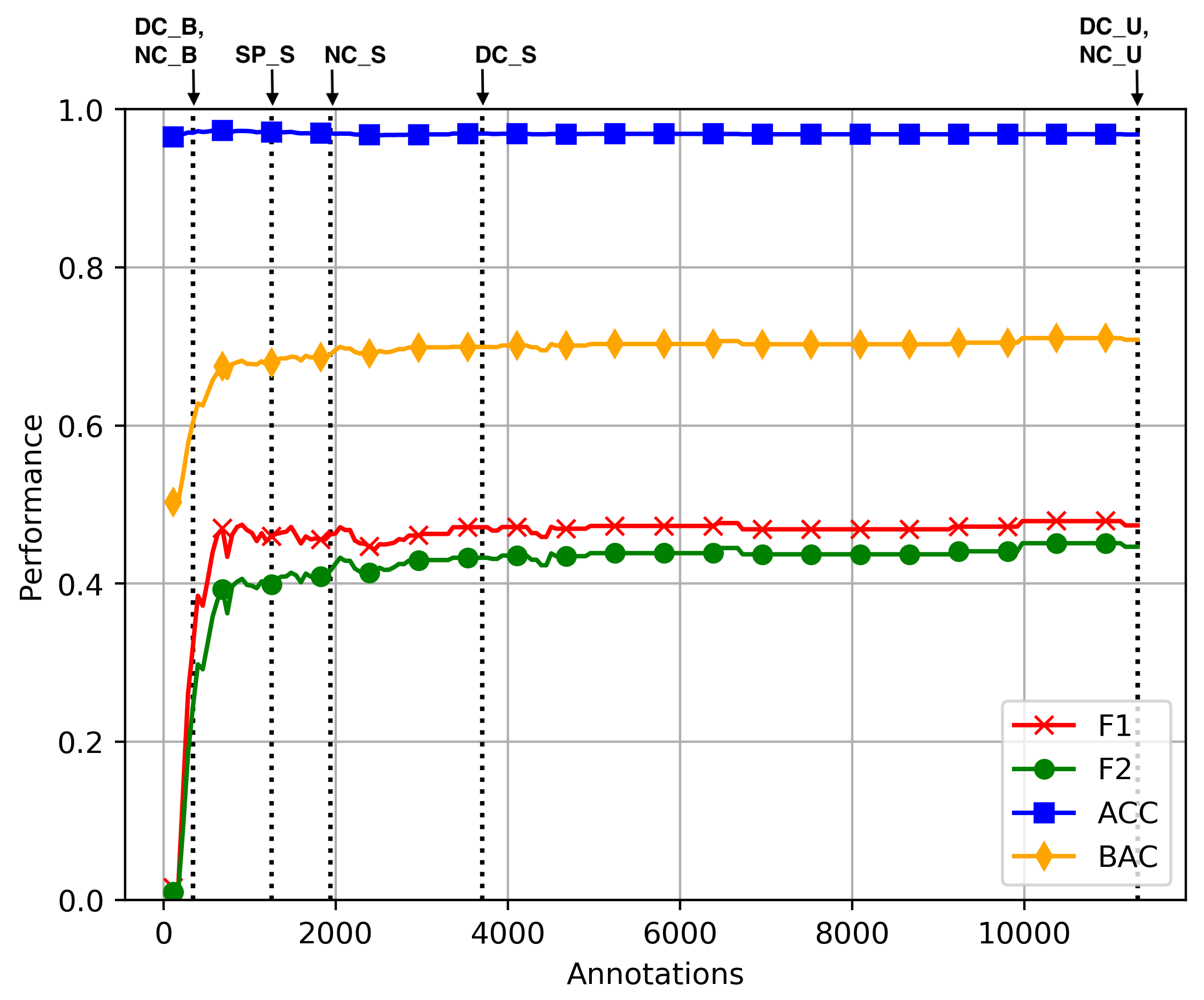}}
\caption{Learning curve for the 20NewsGroups dataset, classifying articles as belonging to the talk-religion category or not. In this plot, $\textrm{DC}_{B}$ and $\textrm{NC}_{B}$ stopped at the same point, so their vertical lines are perfectly on top of each other. The same occurred for $\textrm{DC}_{U}$ and $\textrm{NC}_{U}$.}
\label{fig:LearningCurve}
\end{figure}

Figure~\ref{fig:LearningCurve} shows a representative learning curve for the 20NewsGroups dataset. This learning curve captures common trends that occur at the level of an individual fold or category: the methods that use $U$ fail to stop, the methods that use $B$ stop too early, and the methods that use $S$ stop at reasonable points. The methods that use $S$ are more stable than the other methods. Out of these methods, $\textrm{SP}_{S}$ is the most efficient at saving annotations and producing a high performing model.

%%%%%%%%%%%%%%%%%%%%%%%%%%%%%%%%%%%%%%%%%%%%%%
%%%%%%%%%%%%%%%%%%%%%%%%%%%%%%%%%%%%%%%%%%%%%%

\section{Conclusion} 
\label{sec:Conclusion}

Active learning is used to produce high-performing models while using significantly less labeled training data than when using passive learning. Without an effective stopping method, AL could stop too late, wasting unnecessary data labeling effort, or AL could stop too early, producing a poorly learned model. Several AL stopping methods determine when to stop on the basis of the results of computations performed on a set of examples called the stop set. The choice of the stop set has received relatively less attention than development of the computations that the stopping methods perform on the stop set. In this work, we experiment with several different possible options for a stop set and find that the choice of the stop set can have a significant impact on stopping active learning. It is important to communicate these findings to prevent suboptimal stop sets from being used and suboptimal stopping methods from being used.

Authors of the widely used Stabilizing Predictions method and the Declining Confidence method had originally suggested that representative unbiased sets be used as stop sets for their methods. Recent work used the remaining unlabeled pool and the batch set as stop sets in their new confidence-based methods and also used the remaining unlabeled pool as the stop set when implementing the Stabilizing Predictions method as a baseline for comparison. The remaining unlabeled pool and the batch set are systematically biased as active learning proceeds. 

We find that different stop set choices impact different stopping methods in different ways. When Stabilizing Predictions is implemented with the unlabeled pool instead of an unbiased representative stop set, it becomes more aggressive compared with when it is implemented with an unbiased representative stop set. When confidence-based methods are implemented with the unlabeled pool instead of a large unbiased stop set, they become more conservative, sometimes failing to stop at all. When confidence-based methods are implemented with the batch set, they become less consistent in their behavior and sometimes may become overly conservative or overly aggressive. For both stabilizing predictions and confidence-based methods, we find that the best-performing stop set to use is a random unbiased set. Using one of the biased sets produces a less reliable and lower-performing method. Comparing confidence-based methods with stabilizing predictions methods, our results indicate that when the methods use randomly selected stop sets, the stabilizing predictions method saves many annotations and produces models with higher performance than the confidence-based methods across several publicly available text classification datasets.

%%%%%%%%%%%%%%%%%%%%%%%%%%%%%%%%%%%%%%%%%%%%%%
%%%%%%%%%%%%%%%%%%%%%%%%%%%%%%%%%%%%%%%%%%%%%%

\section*{Acknowledgment} \label{Acknowledgment}

This work was supported in part by The College of New Jersey Support of Scholarly Activities (SOSA) program. The authors acknowledge use of the ELSA high performance computing cluster at The College of New Jersey for conducting the research reported in this paper. This cluster is funded by the National Science Foundation under grant number OAC-1828163. 

\bibliographystyle{./IEEEtran}
\bibliography{paper}

\end{document}